# Packing Moons Inside the Earth

Sunil K. Chebolu

Imagine filling up an earth-sized hollow sphere with the material of the moon. How many moons do we need for this job? In mathematical terms, what is the ratio of the volume of the earth to the volume of the moon? A harder question involves finding the maximum number of solid moons that can be packed inside a hollow earth, as shown in figure 1. If we assume that both the earth and the moon are perfect spheres, then we can address these questions.

### Ancient Greeks and the Lunar Eclipse

A total lunar eclipse occurs when a full moon completely enters the shadow of the earth, as shown in figure 2; it is a beautiful astronomical event. The black region in the picture is known as the umbra and the gray region is called the penumbra.

Observers in the United States witnessed a complete lunar eclipse beginning January 20, 2019, and ending in the early hours of January 21. The sky in Normal, Illinois was clear for this event. I took a picture of this eclipse from my backyard using a Google Pixel 3 camera and Celestron Nexstar 130 mm SLT telescope with a 25 mm eyepiece. Notice the beautiful colors of the sunlight that are reflected from the lunar surface and scattered by the earth's atmosphere in figure 3. The picture was taken when the moon was mostly in the umbra; note the gradual transition in color from the reddish umbral part to the bluish penumbral part.

Assuming that the moon moves at a constant speed, the ancient Greeks argued that the time the moon takes to enter the umbra is proportional to its own diameter, and the time it takes for the moon to cross the umbra is proportional to the diameter of the earth. This simple observation led to the following relationship.

$$\frac{\text{Diameter of the earth}}{\text{Diameter of the moon}} = \frac{\text{Time taken to cross the umbra}}{\text{Time taken to enter the umbra}}.$$

The ancient Greeks used time-keeping devices (water clocks) to estimate that it took 50 minutes for the moon to enter the umbra and 200 minutes for the moon to cross the umbra. Using these approximations, they arrived at the ratio

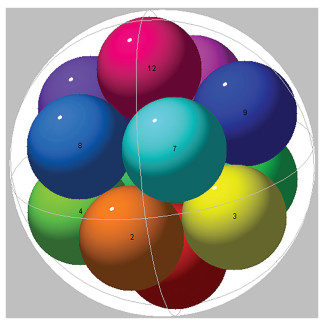

**Figure 1.** Packing moons in the earth.

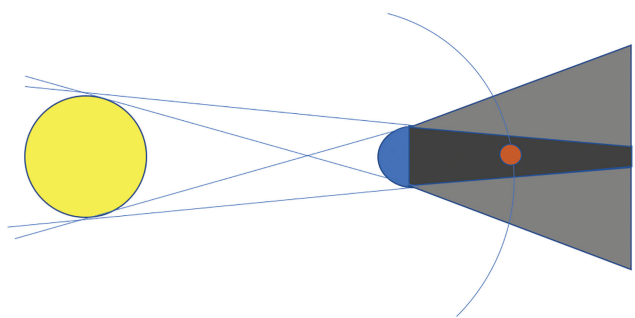

**Figure 2.** The geometry of a lunar eclipse (not drawn to scale).

$$\frac{\text{Diameter of the earth}}{\text{Diameter of the moon}} = \frac{200}{50} = 4.$$

So, they determined that the diameter of the earth is four times the diameter of the moon. If we assume that the earth and the moon are both spheres, then we can use the fact that the volume of a sphere is proportional to the cube of its diameter, so that

$$\frac{\text{Volume of the earth}}{\text{Volume of the moon}} = \left(\frac{\text{Diameter of the earth}}{\text{Diameter of the moon}}\right)^3 = 64.$$

According to this calculation, we would need material from 64 moons to fill up a hollow earth.

Modern measurements put the diameter of the moon at approximately 27.2% of the earth's diameter, so that a better approximation of the volume of the earth to the volume of the moon is $(100/27.2)^3 \approx 50$. We would actually need about 50 moons to fill up a hollow earth.

The error in the ancient Greek estimate arose from lack of precise time-keeping instruments, as well as some incorrect assumptions made in their calculation. As the umbral region is a cone, not a cylinder, the time it takes for the moon to cross the umbra only gives a lower bound on the earth's diameter. Moreover, the described method is valid only when the moon passes through the center of the umbra, which doesn't usually happen. Finally, Kepler's laws imply that the orbital speed of the moon is not constant. It is noteworthy that, despite these shortcomings, the scientific method developed by the ancient Greeks to measure the relative sizes of heavenly bodies based on eclipses is elegant, innovative, and produced relatively accurate answers.

## Packing the Earth with Moons

Note that 50 moons is only an upper bound for the number of *solid* moons that could be packed into a hollow sphere the size of the earth; there will be nonfilled gaps between moons in any packing.

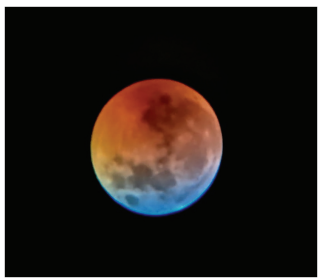

**Figure 3.** Moon during a lunar eclipse.

The challenge of packing moons inside the earth relates to a 400-year-old problem of finding the densest packing of identical spheres (spheres cannot overlap but can touch adjacent spheres) in three-dimensional space known as *the sphere-packing problem*.

The precise problem is to find the arrangement of identical spheres in three-dimensional space so that the spheres fill up the highest possible proportion of the space. This proportion is called the *density* of the sphere packing.

The astronomer Johannes Kepler studied this problem in 1611 and conjectured that the densest way to pack identical spheres in three-dimensional space is the face-centered cubic packing, which is the familiar arrangement of oranges seen in grocery stores as depicted in figure 5.

The left diagram in figure 6 shows a portion of the face-centered cubic packing of unit spheres filling three-dimensional space. The cuboids shown in the diagram tile the entire space when extended indefinitely in all directions. Each cuboid has length and width of two units and contains four eighth-spheres and one half-sphere, as shown on the right in figure 6. The tiling is uniform, so the density of the packing is the same as the density of the partial spheres in one cuboid.

The partial spheres make up a single sphere with volume $(4/3)\pi$.

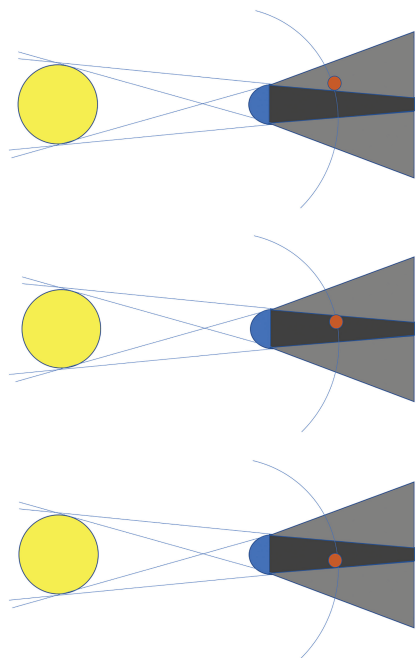

**Figure 4.** A time lapse of a lunar eclipse. Fifty minutes elapse between the first two frames and 200 minutes pass between the first and last frames.

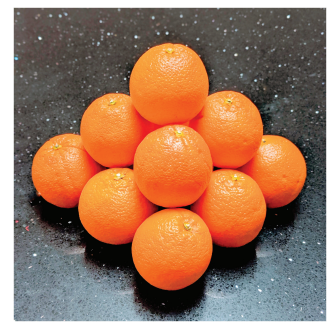

**Figure 5.** The face-centered cubic packing, which is now known to be the densest sphere packing.

**Figure 6.** A portion of the face-centered cubic packing of unit spheres and one cuboid tile that can be uniformly repeated to yield the packing.

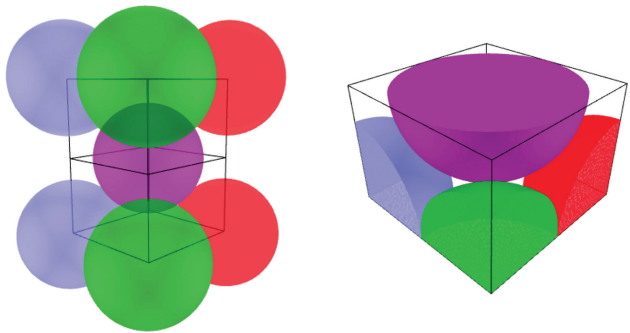

The diagonal connecting opposite corners of two cuboids stacked on top of each other has length 4 as it passes directly through the centers of two half-spheres and one full sphere. Consequently, if the height of one cuboid is $h$, the Pythagorean theorem gives $(2h)^2 + 2^2 + 2^2 = 4^2$. Thus, $h = \sqrt{2}$, and the volume of one cuboid is $4\sqrt{2}$. We can then determine the density of the face-centered cubic packing to be

$$\frac{(4/3)\pi}{4\sqrt{2}} = \frac{\pi}{3\sqrt{2}} = \frac{\pi}{\sqrt{18}}.$$

Despite the plausible and intuitive nature of Kepler's conjecture, the proof turned out to be quite elusive. Gauss provided the first breakthrough in 1831 when he proved that the face-centered cubic packing is the densest packing among all regular packings (those in which the centers of the spheres form a lattice). But showing that this is the densest among all possible packings, including the regular and irregular ones, is much harder. Thomas Hales announced a proof in 1998 that ran 250 pages and involved checking a vast number of cases by extensive computer calculations. The referees were only 99% sure that this proof was correct. Finally, in 2014, Hales produced a formal proof using automated proof checking, settling the conjecture beyond doubt.

Kepler's conjecture implies that in the densest packing, the moons would occupy only $\pi/\sqrt{18} \approx 74.05\%$ of the space inside the earth. Consequently, a better estimate for the maximum number of moons that can be packed inside the earth will be given by

$$(\pi/\sqrt{18}) \cdot \frac{\text{Volume of the earth}}{\text{Volume of the moon}} \approx (\pi/\sqrt{18}) \cdot 50 \approx 37.$$

## Computational Results

Let $a(r)$ denote the maximum number of unit spheres that can be packed inside a sphere of radius $r$. Computing $a(r)$ is hard; the only known values are $a(1) = 1, a(2) = 2, a(3) = 13$, $a(4) = 32$ and $a(5) = 68$. The last two values are experimental (*oeis.org/A084828*). Since the ratio of the diameter of the earth to the diameter of the moon is $100/27.2 \approx 3.67$, which lies between 3 and 4, the maximum number of moons that can be packed inside a hollow earth must be between $a(3) = 13$ and $a(4) = 32$.

We obtain a bit more precision by using Hugo Pfoertner's experimental work, which found packings of 23 spheres with radius 0.275 and 24 spheres with radius 0.271 both inside the unit sphere. Thus, a scaling argument demonstrates that a reasonable estimate for packing moons inside a hollow earth is 23. As the ratio of the volume of the earth to that of the moon is 50, the packing of 23 moons will leave more than half of the space inside the earth empty.

## Further Reading and Final Comments

For more information about the sphere-packing problem, we recommend two articles in *Notices of the AMS*: "A Conceptual Breakthrough in Sphere Packing" by Henry Cohn in 2017 (issue no. 2) or "Kissing Numbers, Sphere Packings, and Some Unexpected Proofs" from 2004 (issue no. 8). We also encourage interested readers to see previous *Math Horizons* articles related to the sphere-packing problem: "Packing Balls in 3, 8, and 24 dimensions" by Sophia Merow from November 2016 and "The Conquest of the Kepler Conjecture" by Dinoj Surendran from April 2001.

To learn more about the computational results we described, we suggest "A Quasi Physical Method for the Equal Sphere Packing Problem," *2011 IEEE 10th International Conference on Trust, Security and Privacy in Computing and Communications*, by WenQi Huang and Liang Yu.

Finally, while we have used sphere-packing in a recreational fashion, we note that sphere-packing results have several practical applications including error-correcting codes, digital communications and crystallography. ●

*Sunil K. Chebolu is a professor of mathematics at Illinois State University. He enjoys sharing his enthusiasm about mathematics with children and college students. His favorite parties are star parties.*

10.1080/10724117.2020.1714329*Sunil K. Chebolu is a professor of mathematics at Illinois State University. He enjoys sharing his enthusiasm about mathematics with children and college students. His favorite parties are star parties.*